\documentclass
[preprint,prl,a4paper,amsfonts,amssymb,showpacs,superscriptaddress,onecolumn]{revtex4}
\usepackage{amssymb}
\usepackage{amsfonts}
\usepackage{amsmath}
\usepackage{graphicx}
\begin{document}

\title{Temperature effects of the magnetic tunnel junctions with periodic grating barrier}
\author{Henan Fang}
\affiliation{College of Electronic and Optical Engineering, Nanjing University of Posts and Telecommunications, Nanjing 210023, China}
\author{Mingwen Xiao\footnote{Email: xmw@nju.edu.cn}}
\affiliation{Department of Physics, Nanjing University, Nanjing 210093, China}
\author{Wenbin Rui}
\affiliation{Department of Physics, Nanjing University, Nanjing 210093, China}
\author{Jun Du}
\affiliation{Department of Physics, Nanjing University, Nanjing 210093, China}
\affiliation{Collaborative Innovation Center of Advanced Microstructures, Nanjing University, Nanjing, 210093, China}
\author{Zhikuo Tao}
\affiliation{College of Electronic and Optical Engineering, Nanjing University of Posts and Telecommunications, Nanjing 210023, China}

\begin{abstract}
We have developed a tunneling theory to describe the temperature dependence of tunneling magnetoresistance (TMR) of the magnetic tunnel junctions (MTJs) with periodic grating barrier. Through the Patterson function approach, the theory can handle easily the influence of the lattice distortion of the barrier on the tunneling process of the electrons. The lattice distortion of the barrier is sensible to the temperature and can be quite easily weakened by the thermal relaxation of the strain, and thus the tunneling process of the electrons gets changed highly with the variation of the temperature of the system. That is just the physical mechanism for the temperature dependence of the TMR. From it, we find that the decrease of TMR with rising temperature is mostly carried by a change in the antiparallel resistance ($R_{AP}$), and the parallel resistance ($R_{P}$) changes so little that it seems roughly constant, if compared to the $R_{AP}$, and that, for the annealed MTJ, the $R_{AP}$ is significantly more sensitive to the strain than the $R_{P}$, and for non-annealed MTJ, both the $R_{P}$ and $R_{AP}$ are not sensitive to the strain. They are both in agreement with the experiments of the MgO-based MTJs. Other relevant properties are also discussed.
\end{abstract}
\maketitle

\section{\textrm{i}. INTRODUCTION}
Magnetic tunnel junctions (MTJs) have received considerable attention for many years. They can be applied to the promising spintronic devices such as high-density magnetic reading head \cite{rf1}. Early experimental studies were limited within the MTJs of amorphous aluminum oxide (Al-O) barriers. In 2001, W. H. Butler et al. \cite{rf2} predicted theoretically that, if MgO single crystal is used to prepare the MTJ barrier, the tunneling magnetoresistance (TMR) can acquire a very high value. The prediction was verified soon by S. S. P. Parkin et al. \cite{rf3} and S. Yuasa et al. \cite{rf4}. Since then, the
MgO-based MTJs have been widely investigated over the last decade \cite{rf5,rf6,rf7,rf8,rf9,rf10,rf11,rf12,rf13}. One of the most important and distinguished properties of MgO-based MTJs is that the parallel resistance $\left( R_{P}\right) $, the antiparallel resistance $\left( R_{AP}\right) $, and the TMR all oscillate with the barrier thickness \cite{rf4,rf10,rf11,rf12,rf13}, which is radically different from the case of Al-O-based MTJs where no such oscillation is found. Those oscillations have already been well interpreted by the spintronic theory developed previously by us \cite{rf14}. The theory is founded on the traditional optical scattering theory \cite{rf15}. Within it, the barrier is treated as a diffraction grating with intralayer periodicity. It is found that the periodic grating can bring strong coherence to the tunneling electrons, the oscillations being a natural result of this coherence. Besides the oscillations, the theory can also explain the puzzle why the TMR is still far away from infinity when the two electrodes are both half-metallic.

Experimentally, there is another important property for MgO-based MTJs, that is, the temperature dependences of the $R_{P}$, $R_{AP}$ and TMR. It is found that, as usual, the TMR will decrease when the temperature of the system increases. However, the decrease of TMR with rising temperature is mostly carried by a change in the $R_{AP}$. The $R_{P}$ changes so little that it seems roughly constant, if compared to the $R_{AP}$ \cite{rf3,rf16,rf17,rf18,rf19,rf20,rf21,rf22,rf23}. Theoretically, the modified version of the magnon excitation model \cite{rf24} is at hand for the mechanism of the above temperature dependence. However, this model can not explain the TMR oscillation on the thickness of MgO barrier. Physically, that is because it dose not include the effect of the periodicity of the single-crystal barrier which plays a key role in the scattering process when the electrons tunnel through the barrier. Based on this reason, we would like to extend our previous theory to interpret the temperature dependences of the $R_{P}$, $R_{AP}$ and TMR of MgO-based MTJs.

As well known, the MgO-based MTJs are fabricated through epitaxial growth. Hence there will be lattice mismatch and interfacial defects between the barrier and its neighbouring layers. Obviously, both of them can cause some lattice distortion to the barrier. The influences of this lattice distortion have been investigated widely by the experiments \cite{rf4,rf25,rf26}. In particular, Ref. [25] discovers that, if the MTJ is annealed, the $R_{AP}$ will increase with raising of strain, which is much more sensitive than the $R_{P}$, and if it is non-annealed, the $R_{AP}$ will unchange with the strain. In addition, Ref. [26] finds that the lattice distortion can modify the band gap of the MgO barrier. Based on those facts, we shall take into account the effect of the lattice distortion of the barrier upon the $R_{P}$, $R_{AP}$ and TMR within the framework of our previous work. Our aim is to interpret theoretically the temperature dependences of the $R_{P}$, $R_{AP}$ and TMR of MgO-based MTJs. As will be seen in the following, this effect can account for the temperature dependences of the $R_{P}$, $R_{AP}$ and TMR of MgO-based MTJs.

\section{\textrm{ii}. METHOD}
To begin with, let us consider a MTJ consisting of a perfect single-crystal barrier. As in Ref. [14], we suppose that the atomic potential of the barrier is $v(\mathbf{r})$, and that the total number of the layers of the barrier is $n$. Then, the periodic potential $U(\mathbf{r})$ of the barrier can be written as
\begin{equation}
U(\mathbf{r})=\sum_{l_{3}=0}^{n-1}\sum_{\mathbf{R}_{h}}v\left( \mathbf{r}-\mathbf{R}_{h}-l_{3}\,\mathbf{a}_{3}\right) ,
\end{equation}
where $\mathbf{R}_{h}$ is a two-dimensional lattice vector of the barrier: $\mathbf{R}_{h}=l_{1}\,\mathbf{a}_{1}+l_{2}\,\mathbf{a}_{2}$, with $\mathbf{a}_{1}$ and $\mathbf{a}_{2}$ being the primitive vectors of the atomic layers, and $l_{1}$ and $l_{2}$ the corresponding integers. The $\mathbf{a}_{3}$ is
the third primitive vector of the barrier, with $l_{3}$ the corresponding integer. Letting $\mathbf{e}_{z}=\mathbf{a}_{1}\times \mathbf{a}_{2}/|\mathbf{a}_{1}\times \mathbf{a}_{2}|$, we shall set $\mathbf{e}_{z}$ point from the upper electrode to the lower one, which is antiparallel to the direction of the tunneling current.

Now, let us consider the effect of the lattice distortion of the barrier. Physically, the periodic potential $U(\mathbf{r})$ of the barrier will be modified by the lattice distortion, as shown by Ref. [26]. In order to elucidate the effect of the distortion on the potential $U(\mathbf{r})$, we would employ the Patterson function approach, which is a standard and very powerful method for studying the diffraction by imperfect crystals \cite{rf15}. Within the framework of two-beam approximation \cite{rf14,rf15}, this leads to that the Fourier transform $v(\mathbf{K}_{h})$ of the atomic potential undergoes a modification as follows,
\begin{equation}\label{vKh0}
  v(\mathbf{K}_{h}) =  \left( 1+2\frac{\sigma }{1-\sigma }\cos \left(
\mathbf{K}_{h}\cdot \mathbf{\alpha }\right) \right) \left( 1-\sigma \right)
v_{0}(\mathbf{K}_{h}),
\end{equation}
where $\mathbf{K}_{h}$ is a planar vector reciprocal to the intralayer lattice vectors $\mathbf{R}_{h}$, $\sigma $ is the defect concentration, $\mathbf{\alpha }$ represents the effect of strain of the barrier \cite{rf15}, and $v_{0}(\mathbf{K}_{h})$ is the Fourier transform of the atomic potential of ideal perfect barrier,
\begin{equation}
  v_{0}(\mathbf{K}_{h}) = \Omega ^{-1}\int d\mathbf{r}\,v(\mathbf{r})e^{-i\mathbf{K}_{h}\cdot \mathbf{r}}.
\end{equation}
Here, $\Omega$ is the volume of the primitive cell of the barrier: $\Omega =(\mathbf{a}_{1}\times \mathbf{a}_{2})\cdot \mathbf{a}_{3}$.

With regard to the strain $\alpha$, Ref. [27] has studied it both experimentally and theoretically on some oxide heterostructures, it is found that, within the low temperature region, the strain decreases linearly with temperature $T$ as follows,
\begin{equation}\label{alpha}
  \alpha = \alpha_{0}\left(1-\frac{T}{T_{c}}\right), \quad T < T_{c},
\end{equation}
where $\mathbf{\alpha }_{0}$ is the strain of the oxide film at zero temperature, and $T_{c}$ is the recovery temperature above which the strain disappears. Generally, $T_{c}$ is around $800\; \mathrm{K}$. As pointed in Ref. [27], this result can be applied to other oxide heterostructures. Therefore, we would like to employee it to handle the strain of MgO barrier. As to the defect concentration $\sigma$, it should be independent on the temperature because the energy to excite defects within a lattice is too high.

Combining the Eqs. (\ref{vKh0}) and (\ref{alpha}) above, we obtain
\begin{equation}\label{vKh}
v(\mathbf{K}_{h}) = \left[ 1+2\frac{\sigma }{1-\sigma }\cos \left(\mathbf{K}_{h}\cdot \mathbf{\alpha }_{0}\left( 1-\frac{T}{T_{c}}\right)\right) \right] \left(1-\sigma \right) v_{0}(\mathbf{K}_{h}).
\end{equation}
This equation builds the relationship between the Fourier transform $v(\mathbf{K}_{h})$ of the atomic potential of realistic imperfect barrier and the temperature $T$.

Now, according to Ref. [14], the transmission coefficient for the channel of the spin-up to spin-up tunneling reads as follows,
\begin{align}
T_{\uparrow \uparrow }(\mathbf{k}) &= \frac{1}{8k_{z}} \Big\{ p_{+}^{z}\mathrm{e}^{i[p_{+}^{z}-(p_{+}^{z})^{\ast }]d}+p_{-}^{z} \mathrm{e}^{i[p_{-}^{z}-(p_{-}^{z})^{\ast }]d}+q_{+}^{z} \mathrm{e}^{i[q_{+}^{z}-(q_{+}^{z})^{\ast }]d}+q_{-}^{z}\mathrm{e}^{i[q_{-}^{z}-(q_{-}^{z})^{\ast }]d}  \notag  \\
&\quad +\Big[p_{+}^{z}\mathrm{e}^{i[p_{+}^{z}-(p_{-}^{z})^{\ast }]d}+p_{-}^{z} \mathrm{e}^{i[p_{-}^{z}-(p_{+}^{z})^{\ast }]d}-q_{+}^{z} \mathrm{e}^{i[q_{+}^{z}-(q_{-}^{z})^{\ast }]d}-q_{-}^{z}\mathrm{e}^{i[q_{-}^{z}-(q_{+}^{z})^{\ast }]d}\Big] + \mathrm{c}.\mathrm{c}. \Big\}
\end{align}
where $\mathbf{k}$ is the incident wave vector of tunneling electrons, and $k_{z}$ its $z$-component, $d$ is the thickness of MgO barrier, and
\begin{subequations}
\begin{align}
p^{z}_{\pm } &= \left[ \mathbf{k}^{2}-\mathbf{k}_{h}^{2}\pm 2m\hbar ^{-2}\,v(\mathbf{K}_{h})\right] ^{1/2}, \\
q^{z}_{\pm } &= \left[ \mathbf{k}^{2}-(\mathbf{k}_{h}+\mathbf{K}_{h})^{2}\pm 2m\hbar ^{-2}\,v(\mathbf{K}_{h})\right]^{1/2}.
\end{align}
\end{subequations}
Here, $\mathbf{k}_{h}$ is the planar component of $\mathbf{k}$. Since $v(\mathbf{K}_{h})$ is a function of $T$ now, the transmission coefficient $T_{\uparrow \uparrow }(\mathbf{k})$ will also be a function of $T$. That is to say, the tunneling process will vary with temperature.

From $T_{\uparrow \uparrow }$, the conductance $G_{\uparrow \uparrow }$ for the channel of the spin-up to spin-up tunneling can be obtained as follows,
\begin{equation}
G_{\uparrow \uparrow } = \frac{e^{2}}{16\pi ^{3}\hbar }\int_{0}^{\pi /2}\text{d}\theta \int_{0}^{2\pi }\text{d}\varphi \,k_{F\uparrow }^{2}\,\sin (2\theta)\,T_{\uparrow \uparrow }\left( k_{F\uparrow },\theta ,\varphi \right),
\end{equation}
where $e$ denotes the electron charge, $\theta$ the angle between $\mathbf{k}$ and $\mathbf{e}_{z}$, $\varphi$ the angle between $\mathbf{k}_{h}$ and $\mathbf{a}_{1}$, and $k_{F\uparrow}$ the Fermi wave vector of the spin-up electrons.
Here, we have ignored the effect of temperature on the Fermi-Dirac distribution of the electrons of ferromagnetic electrodes, which is fairly weak in the present case because $T \leq 400\, \mathrm{K} \ll T_{F}$ where $T_{F} > 10^{4}\; \mathrm{K}$ is the Fermi temperature for either of the electrodes. Since $T_{\uparrow \uparrow }$ is a function of $T$, the above equation shows that $G_{\uparrow \uparrow }$ will depend on the temperature, too.

The other three conductances, $G_{\uparrow \downarrow}$, $G_{\downarrow\uparrow}$, and $G_{\downarrow\downarrow}$, can be obtained similarly. With them, one can get $G_{P} = G_{\uparrow\uparrow} + G_{\downarrow\downarrow}$, $G_{AP} = G_{\uparrow\downarrow} + G_{\downarrow\uparrow}$, $R_{P} = G_{P}^{-1}$, $R_{AP} = G_{AP}^{-1}$, and $\mathrm{TMR} = G_{P}/G_{AP} - 1 = R_{AP}/R_{P} - 1$.

With the same reason as for $G_{\uparrow \uparrow }$, $G_{\uparrow\downarrow}$, $G_{\downarrow\uparrow}$, and $G_{\downarrow\downarrow}$ will also depend on the temperature of the system. Physically, that arises from the fact $v(\mathbf{K}_{h})$ varies with temperature, as shown in Eq. (\ref{vKh}). In a word, the four conductances, $G_{\uparrow\uparrow}$, $G_{\uparrow\downarrow}$, $G_{\downarrow\uparrow}$, and $G_{\downarrow\downarrow}$, as well as the $\mathrm{TMR}$ will all changes with the variation of temperature $T$.

The rest calculations are analogous to the Ref. [14]. The parameters of the ferromagnetic electrodes are chosen as follows: the chemical potential $\mu $ is $11\, \mathrm{eV}$, the half of the exchange splitting $\Delta $ for the ferromagnetic electrodes is $10\, \mathrm{eV}$, and the Fourier transform of the atomic potential of the ideal perfect barrier $v_{0}(\mathbf{K}_{h})$ is set to be $15.3\, \mathrm{eV}$.

\section{\textrm{iii}. RESULTS AND DISCUSSIONS}
As a preparatory step to the temperature effects of the MgO-based MTJs, we shall first study the dependences of $R_{P}$ and $R_{AP}$ on $v(\mathbf{K}_{h})$. The results are shown in Fig. 1 where the thickness of the barrier varies from $1.5\, \mathrm{nm}$ to $3\, \mathrm{nm}$. Obviously, both the $R_{P}$ and $R_{AP}$ oscillate with $v(\mathbf{K}_{h})$. As pointed out in Ref. [14], that originates from the interference among the diffracted waves. In addition, Fig. 1 shows that the amplitude of $R_{AP}$ is much larger than that of $R_{P}$. It can be understood as follows: As stated in Ref. [14], there exist two kinds of integral regions for the transmission coefficients, for the one of them, the transmission coefficients contain oscillating term, for the other, they do not. Only when both $p_{+}^{z}$ and $p_{-}^{z}$ are real or both $q_{+}^{z}$ and $q_{-}^{z}$ are real there can arise oscillating term. For the channel $T_{\downarrow \uparrow }$, the integral regions where the transmission coefficient contains oscillating term is more extensive than the other three channels. It leads to that $R_{AP}$ oscillates more strongly than $R_{P}$. At last, it can be seen from Fig. 1 that the thicker the width of barrier, the shorter the period of the oscillation. Equation (6) indicates that when the width $d$ of the barrier gets thicker, the frequency of $T_{\uparrow \uparrow }$ with respect to $p_{+}^{z}-p_{-}^{z}$ and $q_{+}^{z}-q_{-}^{z}$ will become larger. At the same time, it is easy to know from Eq. (7) that both the $p_{+}^{z}-p_{-}^{z}$ and $q_{+}^{z}-q_{-}^{z}$ are monotonically increasing functions of $v(\mathbf{K}_{h})$. Therefore, the thicker the width $d$ of the barrier, the larger the frequencies of $R_{P}$ and $R_{AP}$ with respect to the variable $v(\mathbf{K}_{h})$.

With those results, the temperature effects of the MgO-based MTJs can be explained as follows. From Eq. (5), it can be found that $v(\mathbf{K}_{h})$ oscillates with temperature $T$. Since both $
R_{P}$ and $R_{AP}$ oscillate with $v(\mathbf{K}_{h})$, as stated above, they will also oscillate with temperature $T$. In addition, the amplitude of $R_{AP}$ will be much stronger than that of $R_{P}$, that is because $R_{AP}$ shows stronger oscillations  with regard to $v(\mathbf{K}_{h})$ than $R_{P}$. This accounts for the physical mechanism of the temperature effects of the MTJ.

In the following, we shall try to use this mechanism to explain in detail the experimental results of the MTJ.

First, we would like to investigate the effect of the strain on $R_{P}$ and $R_{AP}$. The theoretical results are depicted in Fig. 2 where $\mathbf{K}_{h}\cdot\mathbf{\alpha }_{0}$ varies from $\pi/6$ to $\pi/2$, $\sigma  = 0.08$, $T_{c} = 800\, \mathrm{K}$, and $d = 1.5\, \mathrm{nm}$. Figure 2(a) shows the dependence of $R_{P}$ and $R_{AP}$ on the strain when $T = 10\, \mathrm{K}$. Clearly, both $R_{P}$ and $R_{AP}$ oscillate with $\mathbf{K}_{h}\cdot\mathbf{\alpha }_{0}$ but the amplitude of $R_{AP}$ is much larger than that of $R_{P}$. In order to interpret this result, we draw up Fig. 3 to demonstrate the dependence of $v(\mathbf{K}_{h})$ on $\mathbf{K}_{h}\cdot\mathbf{\alpha }_{0}$. It can be seen that $v(\mathbf{K}_{h})$ decreases monotonously from $16.2\, \mathrm{eV}$ to $14.1\, \mathrm{eV}$ when $\mathbf{K}_{h}\cdot\mathbf{\alpha }_{0}$ increases from $\pi/6$ to $\pi/2$. Combining the Figs. 3 and 1, one can easily deduce that the amplitude of $R_{AP}$ is much larger than that of $R_{P}$. As to the annealed MTJ of Ref. [25], it shows that the $R_{AP}$ increases with raising of strain, that can be explained if the strain lies within the range from $\pi/6$ to 1.14 in Fig. 2(a). Of course, if the strain can overcome the region, the $R_{AP}$ would be experimentally expected to decrease or even oscillate with variation of strain. On the other hand, if the MTJ is non-annealed, the barrier of non-annealed MTJ is not well crystallized, the interference arising from the diffraction by the barrier will disappear. Therefore, the $R_{AP}$ will unchange with the strain. In other words, the $R_{AP}$ can not oscillate with the strain. As such, the theoretical results explain the experiments of Ref. [25]: For the annealed MTJ, the $R_{AP}$ is significantly more sensitive to the strain than the $R_{P}$; for non-annealed MTJ, both the $R_{P}$ and $R_{AP}$ are not sensitive to the strain. Figure 2(b) displays the temperature dependence of $R_{P}$ and $R_{AP}$ under different strains. Evidently, both $R_{P}$ and $R_{AP}$ become more sensitive to the temperature when the strain goes larger. That can be easily understood from Eq. (5): The larger the strain is, the more sensitive to the temperature the $v(\mathbf{K}_{h})$ will be.

Secondly, we shall study the effect of $\sigma $ on $R_{P}$ and $R_{AP}$. The results are shown in Fig. 4
where $\sigma $ varies from 0.01 to 0.16, $\mathbf{K}_{h}\cdot
\mathbf{\alpha }_{0} = \pi/3$, $T_{c} = 800\, \mathrm{K}$, and $d = 1.5\, \mathrm{nm}$. Figure 4(a) shows that both $R_{P}$ and $R_{AP}$ are nearly independent on $\sigma $ at $10\,\mathrm{K}$. Physically, that is because $v(\mathbf{K}_{h})$ changes little when $\sigma $ increases from 0.01 to 0.16, as shown in Fig. 5. This can be understood as following: From Eq. (5), we can obtain
 \begin{equation}\label{vKh1}
v(\mathbf{K}_{h}) = v_{0}(\mathbf{K}_{h})+v_{0}(\mathbf{K}_{h})\left[ 2\cos \left(\mathbf{K}_{h}\cdot \mathbf{\alpha }_{0}\left( 1-\frac{T}{T_{c}}\right)\right)-1\right]\sigma.
\end{equation}
Equation (9) shows that there is a linear relationship between $v(\mathbf{K}_{h})$ and $\sigma $. With the present parameters, the slope is very small, therefore, the $v(\mathbf{K}_{h})$ will change little with $\sigma $. Figure 4(b) displays the temperature dependences of $R_{P}$ and $R_{AP}$ with different $\sigma $. Evidently, the larger the $\sigma $, the more sensitive to temperature the $R_{P}$ and $R_{AP}$. It comes from the fact that the larger the $\sigma $, the more sensitive to the temperature the $v(\mathbf{K}_{h})$, as can be easily seen from Eq. (9).

Thirdly, we will discuss the effect of $T_{c}$ on $R_{P}$ and $R_{AP}$. The theoretical results are shown in Fig. 6
where $T_{c}$ varies from $600\, \mathrm{K}$ to $1000\, \mathrm{K}$, $\mathbf{K}_{h}\cdot
\mathbf{\alpha }_{0} = \pi/3$, $\sigma = 0.08$, and $d = 1.5\, \mathrm{nm}$. Figure 6(a) shows the dependence of $R_{P}$ and $R_{AP}$ on $T_{c}$ when temperature is at $10\, \mathrm{K}$. It can be seen that both $R_{P}$ and $R_{AP}$ are nearly independent on $T_{c}$. This can be interpreted from Fig. 7 which shows that $v(\mathbf{K}_{h})$ changes little when $T_{c}$ increases from $600\, \mathrm{K}$ to $1000\, \mathrm{K}$. That is because $T/T_{c}$ is much smaller than 1 when $600\, \mathrm{K} \leq T_{c} \leq 1000\, \mathrm{K}$. From Eq. (5), it means that $v(\mathbf{K}_{h})$ will change little. Figure 6(b) displays the temperature dependence of $R_{P}$ and $R_{AP}$ for different $T_{c}$: The larger the $T_{c}$, the less sensitive to temperature the $R_{P}$ and $R_{AP}$. The result can be easily understood from Eq. (5): The larger the $T_{c}$, the less sensitive to the temperature the $v(\mathbf{K}_{h})$.

Finally, we will compare our theory with experiments. As stated above, the most fundamental feature discovered by the experiments is that the decrease of TMR with rising
temperature is mostly carried by a change in the $R_{AP}$, and the $R_{P}$
changes so little that it seems roughly constant, if compared to the $R_{AP}$%
. In order to reproduce this feature, we draw up Fig. 8 to show the temperature dependences of the $R_{P}$, $R_{AP}$ and TMR where $\mathbf{K}_{h}\cdot
\mathbf{\alpha }_{0} = \pi/3$, $\sigma = 0.08$, $T_{c} = 800\, \mathrm{K}$, and $d = 1.5\, \mathrm{nm}$. With those parameters, the $R_{AP}$ just lies in the dropping region of the oscillation. And because the amplitude of $R_{AP}$ is much larger than that of $R_{P}$, the $R_{P}$
changes so little that it seems roughly constant. It can be seen from Fig. 8 that the theoretical results agree qualitatively well with the experiments \cite {rf3,rf16,rf17,rf18,rf19,rf20,rf21,rf22,rf23}. Here, it should be pointed out that, the experimental results are only within part range of the parameters in the present model. If the whole range is taken into consideration, $R_{P}$ and $R_{AP}$ may decrease, or
increase, or even oscillate with increasing temperature,
which case occurs depends on the varing range of $v(\mathbf{K}_{h})$ when the temperature
changes, as can be easily seen from Fig. 1. This suggests that, if the MgO barrier is replaced by another kind
of material, the $R_{P}$, $R_{AP}$ and TMR may decrease, or increase, or even oscillate with temperature, that is to say, the situation can be quite different from MgO-based MTJs discussed here.

On the other hand, we also calculate the influence of temperature on the TMR
oscillations. The result are shown in Fig. 9 where $\mathbf{K}_{h}\cdot
\mathbf{\alpha }_{0} = \pi/3$, $\sigma = 0.08$, and $T_{c} = 800\, \mathrm{K}$. Figure 9 indicates
that both the amplitude and period decrease weakly with
temperature. This can be understood as follows. When temperature
varies from $10\, \mathrm{K}$ to $300\, \mathrm{K}$, $v(\mathbf{K}_{h})$ will vary correspondingly from
$15.3\, \mathrm{eV}$ to $16\, \mathrm{eV}$. As pointed out in Ref. [14], the amplitude and period of
TMR will both decrease as $v(\mathbf{K}_{h})$ increases. This means that the weak decrease of the amplitude and period roots from the small variation of $v(\mathbf{K}_{h})$, from $15.3\, \mathrm{eV}$ to $16\, \mathrm{eV}$. This theoretical
result is in agreement with the experiments \cite{rf4,rf12}.

\section{\textrm{v}. CONCLUSION}

So far, we have developed a tunneling theory to study the temperature effects of the MTJ with periodic grating barrier. The theory is an extension of our previous work where the barrier is treated as a diffraction grating with intralayer periodicity. Physically, the extension is done mainly through the so-called Patterson function approach. Within the framework of this extension, one can easily take into account the influence of the lattice distortion of the barrier on the tunneling process of the electrons. We find that the distortion can account for the temperature effects of the MTJ with periodic grating barrier.

Theoretically, the distortion of the lattice of the barrier can be described by the defect concentration and the strain, they can both modify highly the scattering potential of the barrier. Although the defect concentration is nearly independent on the temperature, the strain depends strongly upon the temperature of the system. As a result, with the thermal activation of the scattering  potential of the barrier, the tunneling process of the electrons will be highly changed by the temperature of the system, that is just the origination of the temperature effects of the MTJ with periodic grating barrier. With this mechanism, the $R_{P}$, $R_{AP}$ and TMR can all oscillate with the variation of temperature. For a certain concrete range of temperature, the three can occur as increasing, decreasing, or oscillating with temperature. As such, the theory can explain the experiments on the MgO-based MTJs: First, it reproduces the most fundamental feature of the temperature effects: The decrease of TMR with rising temperature is mostly carried by a change in the $R_{AP}$, and the $R_{P}$ changes so little that it seems roughly constant, if compared to the $R_{AP}$. Second, it shows that both the amplitude and period of oscillation of the TMR with regard to the barrier thickness decrease weakly with temperature. And third, it demonstrates that, for the annealed MTJ, the $R_{AP}$ is significantly more sensitive to the strain than the $R_{P}$, and for non-annealed MTJ, both the $R_{P}$ and $R_{AP}$ are not sensitive to the strain.

Recently, Hu and co-workers [28] find an interesting result of the MgO-based MTJs with Co$_{2}$MnSi electrodes. One can easily see from the Fig. (4) of Ref. [28] that the $R_{P}$ oscillates with the temperature, but the $R_{AP}$ dose not. Here, it is worth noting that the situation of Co$_{2}$MnSi electrodes is much distinct from the case considered in this paper because Co$_{2}$MnSi is half-metallic but the present electrodes are conventional. In order to discuss this intriguing property, one needs further to take into account the half-metallic characteristics of the electrodes. We believe that it can be interpreted within the framework of our model. The work is in progress and will be published elsewhere.

\section{ACKNOWLEDGMENTS}
This work is supported by the National Natural Science Foundation of
China (11704197, 61106009, 51471085, 51331004), the State Key Program
for Basic Research of China (2014CB921101), the Nature Science of Foundation
of Jiangsu province (BK20130866), the University Nature Science Research
Project of Jiangsu province (14KJB510020), the Scientific Research
Foundation of Nanjing University of Posts and Communications (NY213025, NY215083, NY217046).

\newpage

\begin{figure}[ht]
\centering
\includegraphics{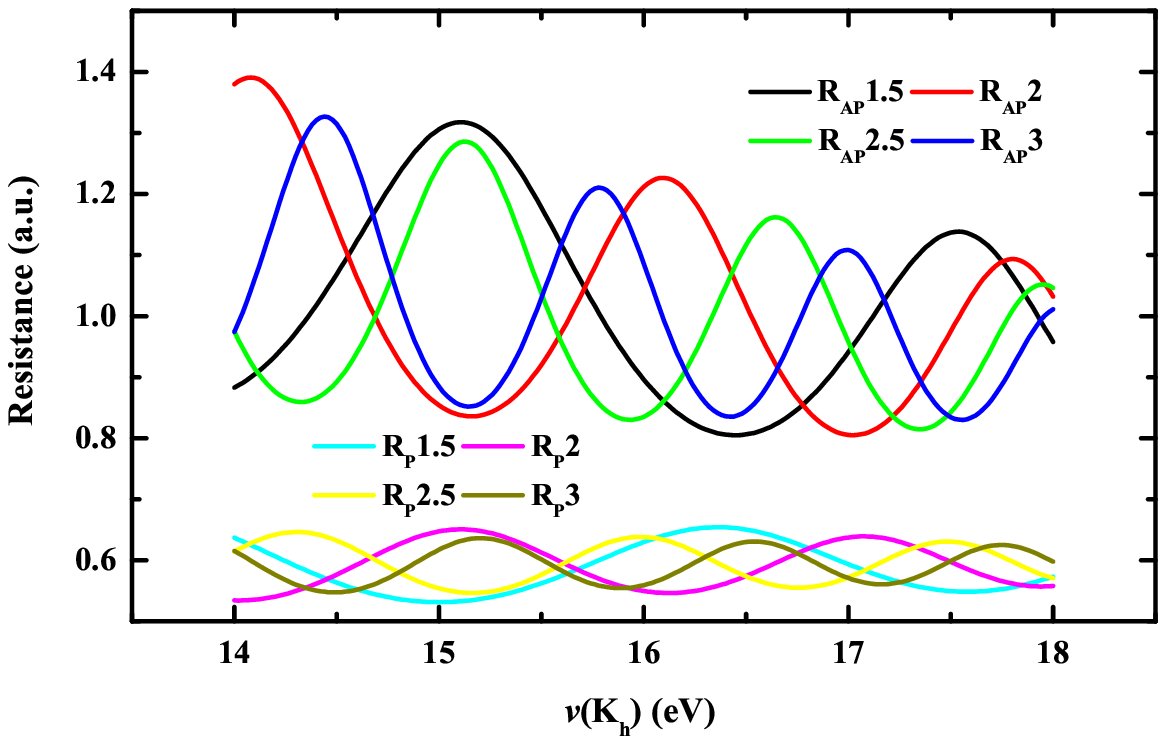}
\caption{$R_{P}$ and $R_{AP}$ as functions of $v(\mathbf{K}_{h}) $ with different barrier thickness $d = 1.5\,\mathrm{nm}$, $2\,\mathrm{nm}$, $2.5\,\mathrm{nm}$, and $3\,\mathrm{nm}$.}
\end{figure}

\begin{figure}[ht]
\centering
\includegraphics{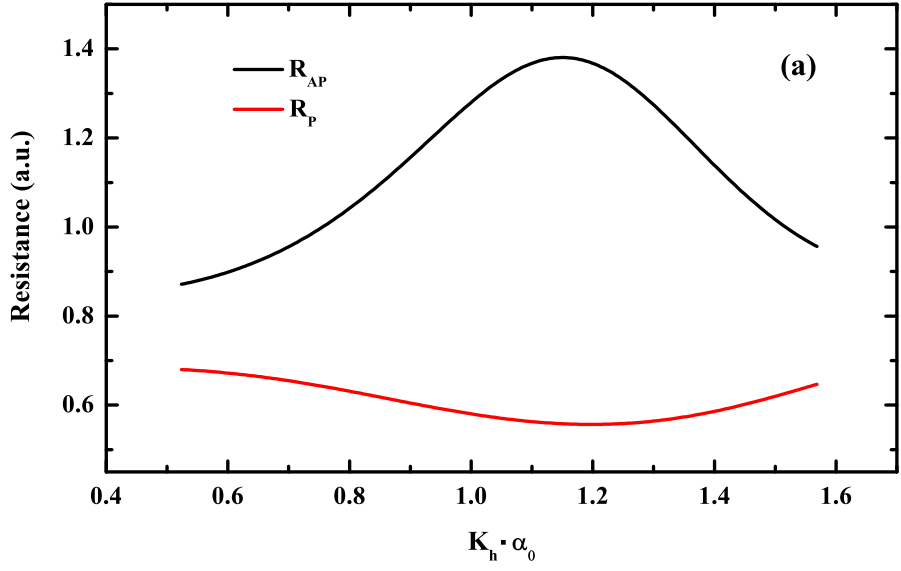}
\includegraphics{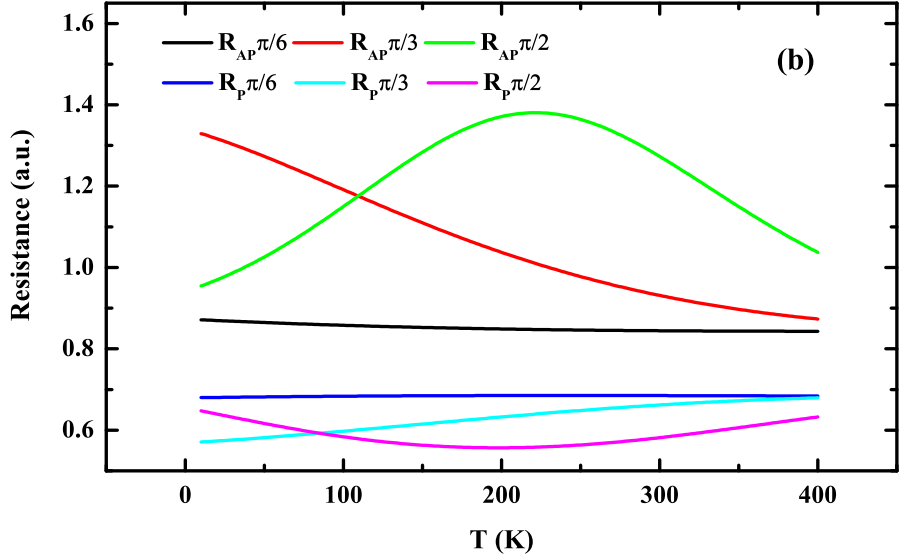}
\caption{(a) $R_{P}$ and $R_{AP}$ as functions of $\mathbf{K}_{h}\cdot
\mathbf{\alpha }_{0}$ at $10\,\mathrm{K}$, (b) $R_{P}$ and $R_{AP}$ as functions of the temperature with different $\mathbf{K}_{h}\cdot
\mathbf{\alpha }_{0} = \pi/6$, $\pi/3$, and $\pi/2$, where $\sigma = 0.08$, $T_{c} = 800\,\mathrm{K}$, and $d = 1.5\,\mathrm{nm}$.}
\end{figure}

\begin{figure}[ht]
\centering
\includegraphics{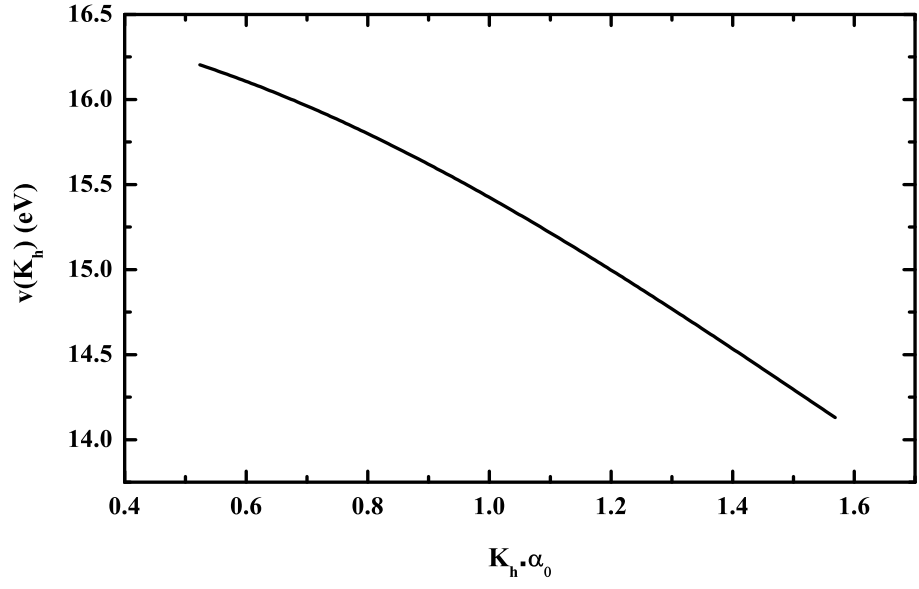}
\caption{At $10\,\mathrm{K}$, $v(\mathbf{K}_{h})$ as functions of $\mathbf{K}_{h}\cdot
\mathbf{\alpha }_{0}$ where $\sigma = 0.08$, $T_{c} = 800\,\mathrm{K}$, and $d = 1.5\,\mathrm{nm}$.}
\end{figure}

\begin{figure}[ht]
\centering
\includegraphics{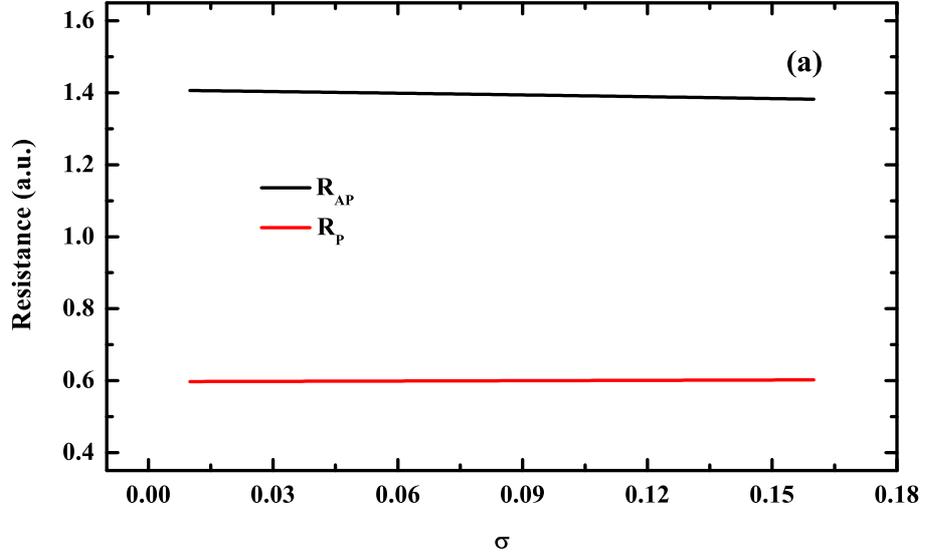}
\includegraphics{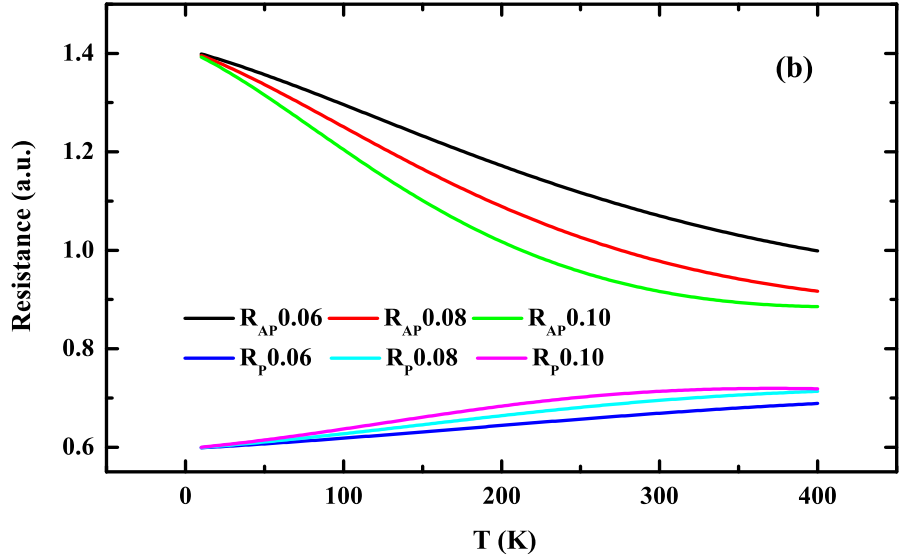}
\caption{(a) $R_{P}$ and $R_{AP}$ as functions of $\sigma $ at $10\,\mathrm{K}$, (b) $R_{P}$ and $R_{AP}$ as functions of the temperature with different $\sigma = 0.06$, $0.08$, and $0.10$, where $\mathbf{K}_{h}\cdot
\mathbf{\alpha }_{0} = \pi/3$, $T_{c} = 800\,\mathrm{K}$, and $d = 1.5\,\mathrm{nm}$.}
\end{figure}

\begin{figure}[ht]
\centering
\includegraphics{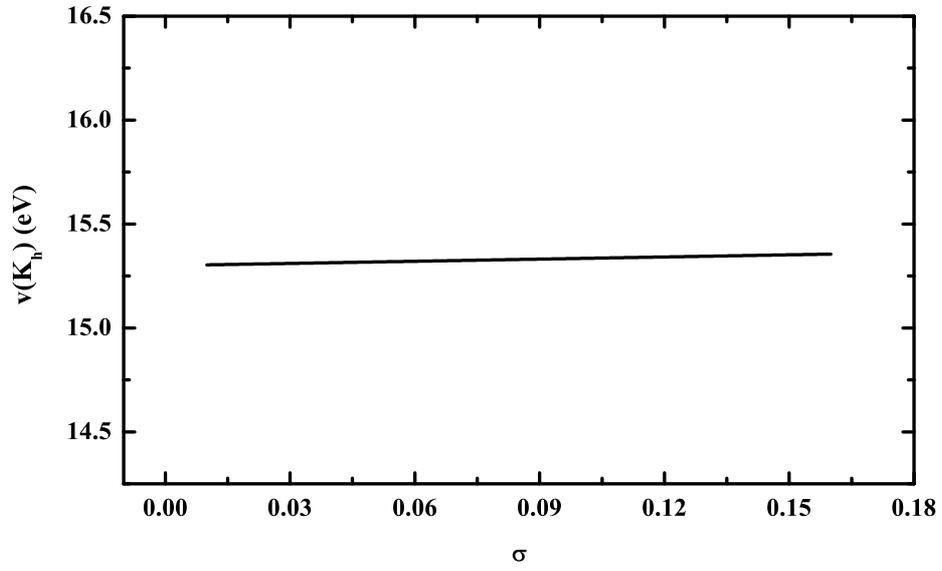}
\caption{At $10\,\mathrm{K}$, $v(\mathbf{K}_{h})$ as functions of $\sigma $ where $\mathbf{K}_{h}\cdot
\mathbf{\alpha }_{0} = \pi/3$, $T_{c} = 800\,\mathrm{K}$, and $d = 1.5\,\mathrm{nm}$.}
\end{figure}

\begin{figure}[ht]
\centering
\includegraphics{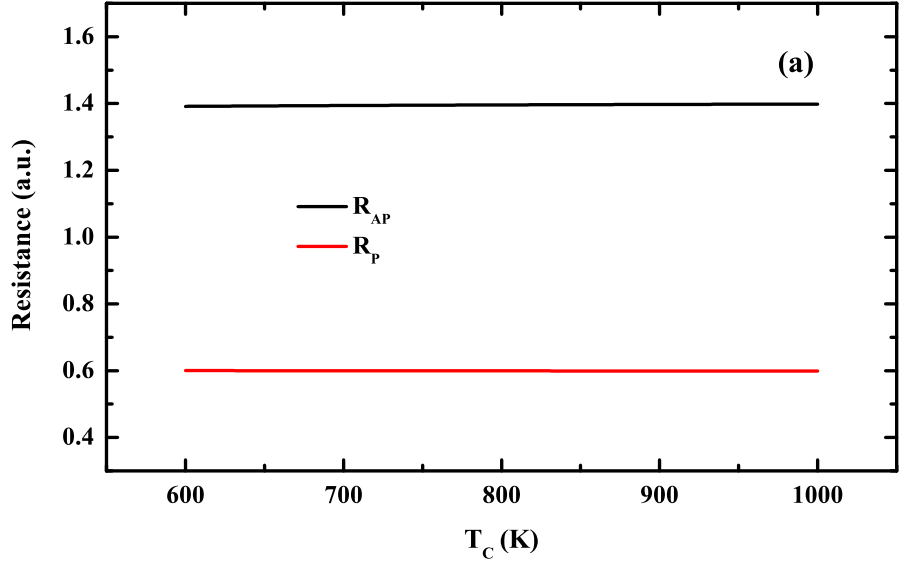}
\includegraphics{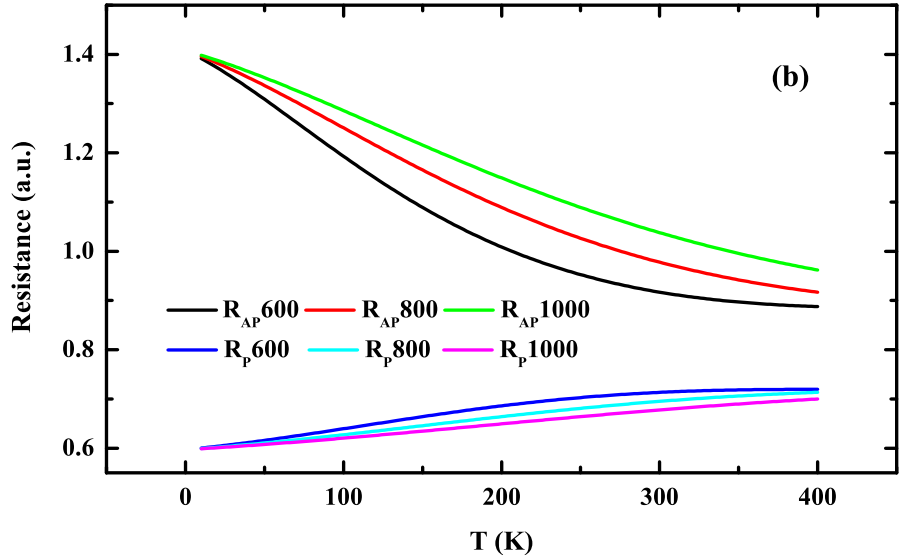}
\caption{(a) $R_{P}$ and $R_{AP}$ as functions of $T_{c}$ at $10\,\mathrm{K}$, (b) $R_{P}$ and $R_{AP}$ as functions of the temperature with different $T_{c} = 600\,\mathrm{K}$, $800\,\mathrm{K}$, and $1000\,\mathrm{K}$, where $\mathbf{K}_{h}\cdot
\mathbf{\alpha }_{0} = \pi/3$, $\sigma = 0.08$, and $d = 1.5\,\mathrm{nm}$.}
\end{figure}

\begin{figure}[ht]
\centering
\includegraphics{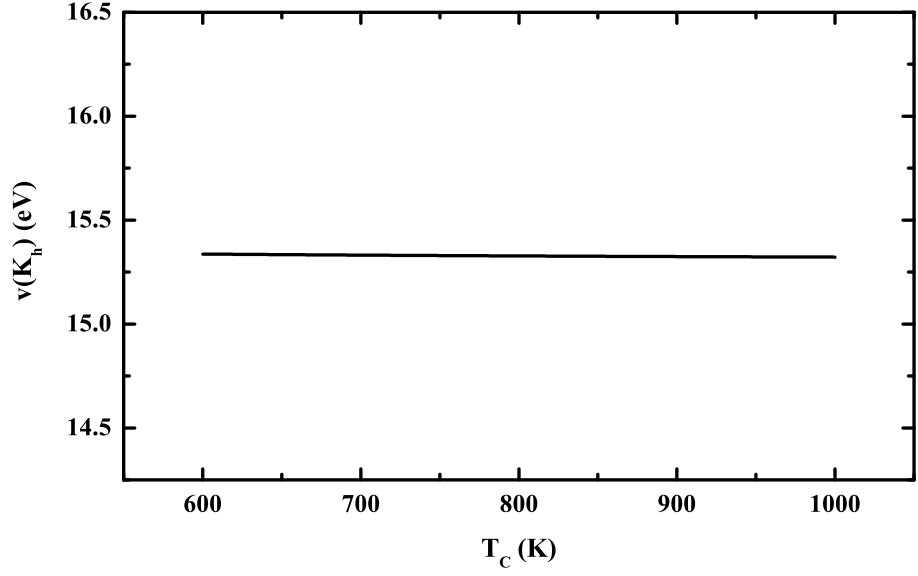}
\caption{At $10\,\mathrm{K}$, $v(\mathbf{K}_{h})$ as functions of $T_{c}$ where $\mathbf{K}_{h}\cdot
\mathbf{\alpha }_{0} = \pi/3$, $\sigma = 0.08$, and $d = 1.5\,\mathrm{nm}$.}
\end{figure}

\begin{figure}[ht]
\centering
\includegraphics{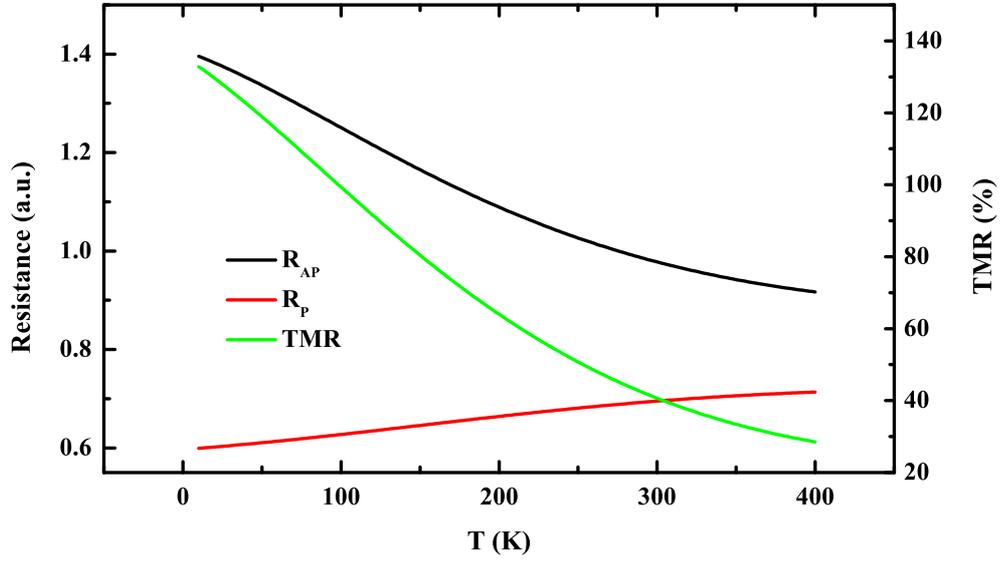}
\caption{$R_{P}$, $R_{AP}$, and TMR as functions of temperature where $\mathbf{K}_{h}\cdot
\mathbf{\alpha }_{0} = \pi/3$, $\sigma = 0.08$, $T_{c} = 800\,\mathrm{K}$, and $d = 1.5\,\mathrm{nm}$.}
\end{figure}

\begin{figure}[ht]
\centering
\includegraphics{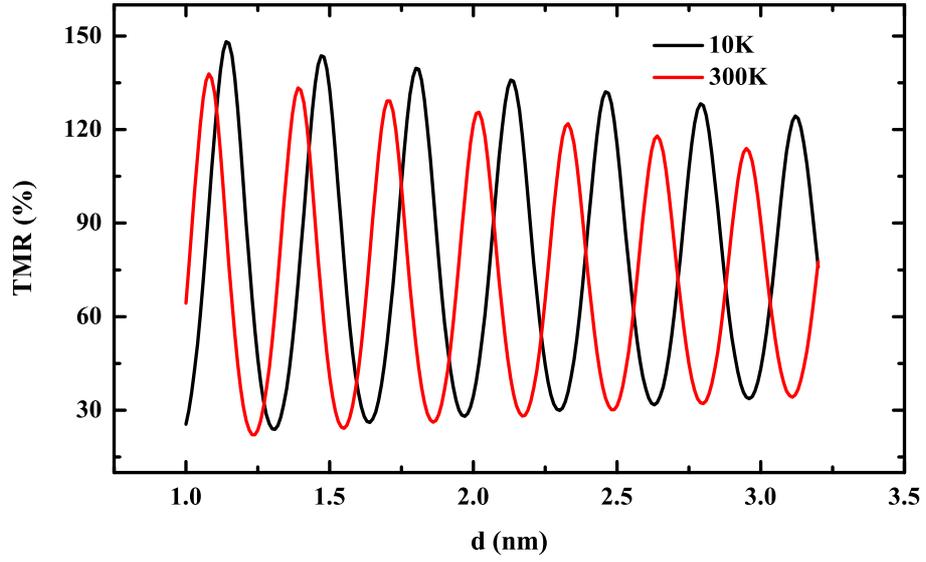}
\caption{TMR as functions of barrier thickness $d$ at $10\,\mathrm{K}$ and $300\,\mathrm{K}$ where $\mathbf{K}_{h}\cdot
\mathbf{\alpha }_{0} = \pi/3$, $\sigma = 0.08$, and $T_{c} = 800\,\mathrm{K}$.}
\end{figure}

\end{document}